\documentclass[reprint, superscriptaddress, amsmath, amssymb, aps, pra]{revtex4-2}

\usepackage{graphicx}
\usepackage{dcolumn}
\usepackage{xcolor}
\usepackage{bm}
\usepackage{bbm}
\usepackage{braket}
\usepackage{hyperref}
\usepackage{comment}
\usepackage{ragged2e}
\usepackage{algpseudocode}
\usepackage{placeins}
\usepackage{amssymb}

\usepackage[capitalize]{cleveref}
\crefname{section}{Sec.}{Secs.}
\Crefname{section}{Section}{Sections}

\crefrangelabelformat{section}{#3#1#4--#5\crefstripprefix{#1}{#2}#6}
\crefrangelabelformat{figure}{#3#1#4--#5\crefstripprefix{#1}{#2}#6}
\crefrangelabelformat{equation}{(#3#1#4--#5\crefstripprefix{#1}{#2}#6)}

\crefmultiformat{equation}%
{\edef\crefstripprefixinfo{#1}Eqs.~(#2#1#3}%
{,#2\crefstripprefix{\crefstripprefixinfo}{#1}#3)}%
{,#2\crefstripprefix{\crefstripprefixinfo}{#1}#3}%
{,#2\crefstripprefix{\crefstripprefixinfo}{#1}#3)}
\definecolor{darkblue}{rgb}{0,0.3,0.7}
\hypersetup{
    colorlinks=true,
    linkcolor=darkblue,
    filecolor=blue,
    citecolor=darkblue,
    urlcolor=darkblue,
}

\DeclareMathOperator{\Tr}{Tr}
\def\bx{\mathbf{x}}

\def\bcD{{\bm{\mathcal{D}}}}

\def\cZ{\mathcal{Z}}
\def\cF{\mathcal{F}}
\def\cR{\mathcal{R}}

\def\cQ{\mathcal{Q}}
\def\cE{\mathcal{E}}
\def\cP{\mathcal{P}}

\def\id{\mathbbm{1}}
\def\imag{\text{i}}

\usepackage[T1]{fontenc}
\usepackage[utf8]{inputenc}

\begin{document}

\preprint{APS/123-QED}

\title{In situ quantum verification of polarization-stabilized optical channels}  

\author{Matthew~L. Stevens}
\email{steve515@purdue.edu}
\affiliation{Elmore Family School of Electrical and Computer Engineering and Purdue Quantum Science and Engineering Institute, Purdue University, West Lafayette, Indiana 47907, USA}
\affiliation{School of Electrical, Computer, and Energy Engineering and Research Technology Office, Arizona State University, Tempe, Arizona 85287, USA}
\author{Noah~I. Wasserbeck}
\affiliation{School of Electrical, Computer, and Energy Engineering and Research Technology Office, Arizona State University, Tempe, Arizona 85287, USA}
\author{Zachary Goisman}
\affiliation{Elmore Family School of Electrical and Computer Engineering and Purdue Quantum Science and Engineering Institute, Purdue University, West Lafayette, Indiana 47907, USA}
\affiliation{School of Electrical, Computer, and Energy Engineering and Research Technology Office, Arizona State University, Tempe, Arizona 85287, USA}
\author{Arefur Rahman}
\affiliation{Elmore Family School of Electrical and Computer Engineering and Purdue Quantum Science and Engineering Institute, Purdue University, West Lafayette, Indiana 47907, USA}
\affiliation{School of Electrical, Computer, and Energy Engineering and Research Technology Office, Arizona State University, Tempe, Arizona 85287, USA}
\author{John Michael Record}
\affiliation{School of Electrical, Computer, and Energy Engineering and Research Technology Office, Arizona State University, Tempe, Arizona 85287, USA}
\author{Taman Truong}
\affiliation{School of Electrical, Computer, and Energy Engineering and Research Technology Office, Arizona State University, Tempe, Arizona 85287, USA}
\author{Ariq Haqq}
\affiliation{School of Electrical, Computer, and Energy Engineering and Research Technology Office, Arizona State University, Tempe, Arizona 85287, USA}
\author{Muneer Alshowkan}
\affiliation{Quantum Information Science Section, Oak Ridge National Laboratory, Oak Ridge, Tennessee 37831, USA}
\author{Brian~T. Kirby}
\affiliation{DEVCOM US Army Research Laboratory, Adelphi, Maryland 20783, USA}
\affiliation{Tulane University, New Orleans, LA 70118, USA}
\author{Nils T. Otterstrom}
\affiliation{Microsystems Engineering, Science, and Applications, Sandia National Laboratories, Albuquerque, New Mexico 87185, USA}
\author{Joseph~M. Lukens}
\email{jlukens@purdue.edu}
\affiliation{Elmore Family School of Electrical and Computer Engineering and Purdue Quantum Science and Engineering Institute, Purdue University, West Lafayette, Indiana 47907, USA}
\affiliation{School of Electrical, Computer, and Energy Engineering and Research Technology Office, Arizona State University, Tempe, Arizona 85287, USA}
\affiliation{Quantum Information Science Section, Oak Ridge National Laboratory, Oak Ridge, Tennessee 37831, USA}

\date{\today}

\begin{abstract}
The active stabilization of polarization channels is a task of growing importance as quantum networks move to deployed demonstrations over existing fiber infrastructure. However, the uniquely strict requirements for high-fidelity qubit transmission  complicate the extent to which classical solutions may apply to future quantum networks, particularly in terms of recognizing noise sources present in low-flux, nonunitary channels.
Here we introduce a novel in situ benchmarking approach that augments a classical polarization tracking system, limited to unitary correction, with simultaneously transmitted quantum light for ancilla-assisted process tomography of the full quantum map. Implemented in a local-area quantum network, our method uses the reconstructed map both to validate the classical compensation and to expose noise sources it fails to capture. A sliding measurement window that continuously updates the estimated quantum process further increases sensitivity to rapid channel fluctuations. Our results should unlock new opportunities for in situ channel characterization in quantum--classical coexistence networks.
\end{abstract}
\maketitle

\section{Introduction}
Ancilla-assisted process tomography (AAPT) is a powerful tool leveraging quantum correlations to  characterize some unknown quantum process~\cite{Altepeter2003AAPT,lie2023faithfulness}. Recently applied to a deployed fiber channel~\cite{arefur2024AAPT}, AAPT enlists an auxiliary system local to Alice to extract the process from Alice to Bob purely from the joint output state, thereby permitting characterization of quantum channels with in situ entanglement resources. Yet, because AAPT is fundamentally a tomographic tool, it can \emph{identify} noise in a channel but cannot by itself \emph{mitigate} such noise---an omnipresent need in quantum information science in general, and polarization encoding in particular, due to time-dependent birefringence in optical fibers.

For this reason, a large body of quantum communications experiments have adopted concurrent polarization tracking and compensation techniques to preserve high-fidelity qubit transmission in the face of environmental fluctuations.
Whether based on time \cite{treiber2009-8,craddock2024automated17,kucera2024quantum} or frequency multiplexing~\cite{xavier2008polarization14,xavier2009qkd15,pan2023cvqkdtracking,Chapman2024} of a parallel high-flux classical probe, direct measurement of quantum bit error rate (QBER)~\cite{Neumann2022, shi2021polarizationcomp, peranic2023polarization16}, or combinations thereof~\cite{peranic2023polarization16}, the basic principle is the same: adjust electrical or mechanical waveplates in real time to stabilize the polarization transformation measured by the tracking system. Such a feedback system amounts to a controllable unitary rotation $U_\mathrm{feedback}$ in the polarization-qubit Hilbert space; hence, if the channel itself is described by a unitary matrix $U_\mathrm{channel}$ that is slowly varying on the timescales of the feedback system, the feedback condition $U_\mathrm{feedback}=U_\mathrm{channel}^\dagger$ yields a perfect identity channel $U_\mathrm{feedback}U_\mathrm{channel} = \mathbbm{1}$.

Yet, while the slowly varying unitary approximation to a fiber-optic channel captures the dominant sources of impairments in many practical scenarios, it overlooks many others, such as polarization-mode dispersion~\cite{Gordon2000, Brodsky2006, Antonelli2011}, polarization-dependent loss (PDL)~\cite{Bongioanni2010, Jones2018, Kirby2019}, and background light (e.g., Raman noise)~\cite{Stolen1984, Peters2009, chapman2023coexistent, Thomas2023}. Consequently, even an ideal polarization compensation system may fail to yield an ideal quantum channel for two main reasons: (i)~the nonunitary portion of the polarization transformation cannot be corrected by waveplates alone, and (ii)~the low-flux quantum signals will often operate at a significantly lower signal-to-noise ratio than the classical tracking fields. Thus, independent quantum validation of any polarization compensation system is critical to establishing its utility within a specific channel. To our knowledge, only one example has fully characterized a polarization tracking channel with an in situ quantum resource~\cite{Chapman2024}---and only for a fiber spool---leaving approaches that fully characterize the quantum channel and operate at precisely the quantum conditions of interest largely unexplored.

In this work, we propose and systematically investigate an AAPT-based system for real-time validation of a spectrally multiplexed polarization compensation system in a deployed three-node quantum local-area network.
AAPT takes 163~s per fully independent tomography, with a ``sliding window'' modification that increases the update rate to every 10~s by leveraging overlapping datasets. 
Our approach succeeds in confirming the success of the classical gradient-descent-based compensation system while also revealing nonunitary errors to which the unitary classical system is blind. Extending the long pedigree of wavelength-division multiplexing in quantum--classical coexistence~\cite{Townsend1997b, Peters2009DWDMQKD, Eraerds2010, Patel2012, Eriksson2019CVQKD, chapman2023coexistent}, our concurrent measurement of distinct metrics (reference performance calculated classically and quantum performance computed via AAPT) offers direct quantum validation with minimal intrusion into channel use.


\section{Background}
A common approach to mitigate noise within a quantum link involves frequency or time multiplexing the quantum signal with a classical reference to monitor the system in real time ~\cite{kucera2024quantum,treiber2009-8,xavier2008polarization14,xavier2009qkd15,peranic2023polarization16,craddock2024automated17, Chapman2024,pan2023cvqkdtracking}. The results of classical feedback have also been used to infer the full quantum process~\cite{kucera2024quantum,ndagano2016processtomographyquantumchannels}. For a unitary reference channel $\cE_\cR$ operating on a $d$-dimensional density matrix $\rho$ as
\begin{equation}
\label{eq:channelU}
\mathcal{E}_\cR(\rho)=U\rho U^\dagger,
\end{equation}
the process fidelity with respect to an ideal target unitary $T$ can be written as~\cite{Raginsky2001, nielsen2002fidelity}
\begin{equation}
\label{eq:fidU}
\cF_\cR = \frac{1}{d^2}\left|{\Tr T^\dagger U}\right|^2.
\end{equation}
We use the subscript $\cR$ for ``reference fidelity'' to distinguish it from the full ``quantum fidelity'' introduced later.
In our case of polarization-qubit transmission in optical fiber, $d=2$, and $U$ can be expressed generically as
\begin{equation}
\label{eq:IdentityAndU}
U = 
\begin{bmatrix}
\cos\frac{\theta}{2} & -e^{\imag \lambda} \sin\frac{\theta}{2} \\
e^{\imag \psi} \sin\frac{\theta}{2} & e^{\imag (\psi + \lambda)} \cos\frac{\theta}{2}
\end{bmatrix}
\end{equation}
for $\theta\in[0,\pi]$ and $\psi,\lambda\in[0,2\pi)$ \cite{Nielsen2000}.

The unitary $U$ can be estimated  directly from classical power measurements. Specifically, 
assuming the launched classical signal is switched between horizontal $\ket{H}$ and diagonal $\ket{D}=\frac{1}{\sqrt{2}}(\ket{H}+\ket{V})$ states and measured after the channel in $\ket{H}$, $\ket{V}$, $\ket{D}$, and $\ket{A}=\frac{1}{\sqrt{2}}(\ket{H}-\ket{V})$ projections, the parameters in \cref{eq:IdentityAndU} can be computed as follows:
\begin{equation}
\begin{split}
\theta & = 2\cos^{-1}\sqrt{f_{H|H}}  \\ 
\psi & = \cos^{-1}\left(\frac{2f_{D|H} - 1}
    {2\cos\frac{\theta}{2}\sin\frac{\theta}{2}}\right) \label{eq:angles} \\
\lambda & = \cos^{-1}\left(\frac{1- 2f_{H|D} } 
    {2\cos\frac{\theta}{2}\sin\frac{\theta}{2}}\right),
\end{split}
\end{equation}
where the fractional powers are defined as
\begin{equation}
\begin{split}
\label{eq:fractions}
f_{H|H} & = \frac{\cP(H|H)}{\cP(H|H)+\cP(V|H)} \\
f_{D|H} & = \frac{\cP(D|H)}{\cP(D|H)+\cP(A|H)} \\
f_{H|D} & = \frac{\cP(H|D)}{\cP(H|D)+\cP(V|D)}.
\end{split}
\end{equation}
Here $\cP(Y|X)$ denotes the power measured in the output projection $Y\in\{H,V,D,A\}$ given input state $X\in\{H,D\}$. 

\begin{figure*}[bt!]
\includegraphics[width=\textwidth]{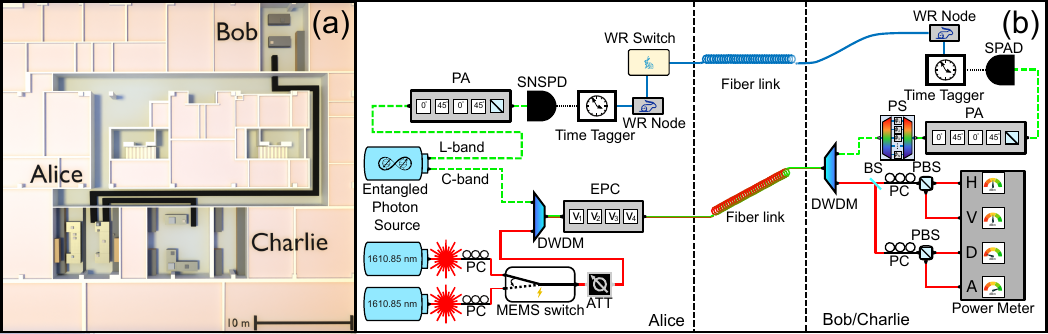}
    \caption{(a) Fiber lightpaths between labs in the subbasement of the MSEE building at Purdue University. (b) Schematic of the Alice--Bob link (duplicated for Alice--Charlie) including the parallel fiber link for classical communications. WR: White Rabbit; SNSPD: superconducting nanowire single-photon detector; SPAD: single-photon avalanche diode; PA: polarization analyzer; DWDM: dense wavelength division multiplexing filter; ATT: attenuator; PC: manual polarization controller; EPC: electronic polarization controller; PS: pulse shaper; BS: fiber beamsplitter; PBS: fiber polarizing beamsplitter.}
    \label{fig:Setup}
\end{figure*}

\Crefrange{eq:IdentityAndU}{eq:fractions} fully characterize any unknown unitary polarization channel. In practice, however, the quantum process will display nonunitary features as well, in which case Eq.~\eqref{eq:channelU} can be generalized to the noisy quantum channel
\begin{equation}
\label{eq:channelQ}
\mathcal{E}_\cQ(\rho) = \sum_{k=1}^R A_k \rho A_k^\dagger,
\end{equation}
where $R\leq d^2$ denotes the Choi rank and the Kraus operators $\{A_k\}$ satisfy $\sum_k A_k^\dagger A_k = \mathbbm{1}$; i.e., $\mathcal{E}_\cQ$ is completely positive and trace-preserving (CPTP)~\cite{Wilde2017}. (Our method can be extended to trace-decreasing maps via models like that of Ref.~\cite{Bongioanni2010}, but for simplicity of presentation---and given the negligible PDL in our physical system---we retain the CPTP constraint here.) For this channel, the process fidelity with respect to the unitary target $T$ can be written as \cite{johnston2011choi}
\begin{equation}
\label{eq:krausFid}
\cF_\cQ = \frac{1}{d^2}\sum_{k=1}^R \left|\Tr T^\dagger A_k \right|^2,
\end{equation}
as compared to the single-term Eq.~\eqref{eq:fidU}.
We experimentally estimate this full quantum channel $\cE_\cQ$ via Bayesian AAPT. Details on our approach to Bayesian process tomography can be found in Ref.~\cite{chapman2023coexistent}, with the extension to AAPT following in Ref.~\cite{arefur2024AAPT}. Here we briefly outline the basic features. Beginning with a known two-qubit input state $\rho_{\alpha\beta}$, the ancilla photon $\alpha$ is kept locally while the probe photon $\beta$ traverses the unknown channel of interest, after which a projective measurement is performed on both photons in some pure product state $\ket{\varphi_s}$. Parameterizing the Kraus operators by a vector $\bx$, the probability associated with this measurement is
\begin{equation}
\label{eq:ps}
\begin{split}
p_s(\bx) & =\braket{\varphi_s|\left(\id \otimes \cE_\cQ\right)(\rho_{\alpha\beta})|\varphi_s} \\
 &=\sum_{k=1}^{R} \braket{\varphi_s|[\id\otimes A_k(\bx)]\rho_{\alpha\beta}[\id\otimes A_k^\dagger(\bx)]|\varphi_s}.
\end{split}
\end{equation}

These probabilities and empirical counts $\bcD=\{N_s\}$ comprise a Poissonian likelihood
\begin{equation}
\label{eq:LL}
L_\bcD(\bx) =\prod_{s=1}^S e^{-K(\bx)p_s(\bx)}[p_s(\bx)]^{N_s},
\end{equation}
where $K(\bx)$ is an unknown flux parameter. Following Ref.~\cite{Kukulski2021}, a standard normal prior distribution
\begin{equation}
\label{eq:prior}
\pi_0(\bx) =\prod_{i=1}^{2d^4+1} \frac{1}{\sqrt{2\pi}}e^{-\frac{1}{2}x_i^2},
\end{equation}
in which the $R=d^2$ Kraus operators are each constructed by $2d^2$ real parameters and normalized to a CPTP map, covers the space of arbitrary quantum processes uniformly under the Lebesgue measure. The final parameter $x_{2d^4+1}$ specifies the flux $K(\bx) = K_0(1+\sigma_0 x_{2d^4+1})$, where we take $\sigma_0=0.1$ and define $K_0$ from the measured counts~\cite{Simmerman2020, Lu2022b}.

Formally, the posterior distribution is then
\begin{equation}
\label{eq:posterior}
\pi(\bx) =\frac{1}{\cZ}L_\bcD(\bx)\pi_0(\bx),
\end{equation}
with $\cZ$ an undetermined normalization constant. We sample from $\pi(\bx)$ leveraging an efficient preconditioned Crank--Nicolson (pCN) Markov chain Monte Carlo (MCMC) Bayesian tomography workflow---first proposed in Ref.~\cite{Lukens2020b}, extended to AAPT in Ref.~\cite{arefur2024AAPT}, and parallelized in Ref.~\cite{Nguyen2025}---and draw $N$ random samples $\{\bx^1,...,\bx^N\}$ from the posterior $\pi(\bx)$, each corresponding to a full set of Kraus operators, i.e., $\cE_\cQ^i=\{A_1(\bx^i),...,A_R(\bx^i)\}$. Because the Kraus operators are not unique, we express each channel as a Choi matrix~\cite{Wilde2017} defined with respect to the computational basis $\{\ket{1},...,\ket{d}\}$ as
\begin{equation}
\label{eq:choi}
\begin{split}
\Phi_\cQ(\bx^i) &= \frac{1}{d} \sum_{m=1}^d\sum_{n=1}^{d} \ket{m}\bra{n} \otimes \cE_\cQ^i(\ket{m}\bra{n}) \\
&=\frac{1}{d} \sum_{m=1}^{d}\sum_{n=1}^{d}\sum_{k=1}^{d^2} \ket{m}\bra{n} \otimes \left[A_k(\bx^i)\ket{m}\bra{n}A_k^\dagger(\bx^i)\right],
\end{split}
\end{equation}
from which the Bayesian mean estimator $\hat{\Phi}_\cQ$ is
\begin{equation}
\label{eq:choiBayes}
\hat{\Phi}_\cQ = \frac{1}{N}\sum_{i=1}^N\Phi_\cQ(\bx^i).
\end{equation}

In light of the preceding theory, the experiments that follow below can be viewed as the concurrent implementation of two optical probes based on distinct models of the deployed quantum channel: (i) a classical, bright, and fast polarization tracking system based on a unitary description of the fiber link [\crefrange{eq:channelU}{eq:fractions}]; and (ii) a quantum, low-flux, and slower AAPT-based monitor for independent validation [\crefrange{eq:channelQ}{eq:choiBayes}].

\section{Experiment}
\label{sec:experiment}
Figure~\ref{fig:Setup}(a) shows the quantum local area network comprising three optics laboratories in the subbasement of the Materials and Electrical Engineering (MSEE) Building at Purdue University. An approximately 86~m length of two SMF-28 optical fibers connects Alice to Bob via a dropped ceiling, while Alice and Charlie are connected by $\sim$60~m SMF-28 lightpaths. For all tests, the classical and quantum light sources reside in Alice and are multiplexed in the same optical fiber strand en route to either Bob or Charlie.

A schematic of the key experimental components follows in Fig.~\ref{fig:Setup}(b).
Classical tracking is based on a simple polarimeter at the receiver, which is first calibrated in a self-consistent, self-contained fashion. Specifically, the state $\ket{H}$ at the input to the 50/50 fiber beamsplitter is defined such that the first two-port power ratios satisfy $f_{H|H}=1$ and $f_{V|H}=0$ [\cref{eq:fractions}]. Then, keeping this input state fixed, the manual polarization controller preceding the $D/A$ polarizing beamsplitter is adjusted to achieve $f_{D|H}=f_{A|H}=0.5$. Under these settings, the two mutually unbiased states required for polarization tracking can be defined operationally as those which achieve $(f_{H|H},f_{V|H},f_{D|H},f_{A|H})=(1,0,0.5,0.5)$ and $(f_{H|D},f_{V|D},f_{D|D},f_{A|D})=(0.5,0.5,1,0)$.

Two lasers tuned to 1610.85 nm and attenuated to approximately $-49$~dBm received power at the polarimeter (to minimize crosstalk with the quantum channel while remaining comfortably above the $-80$~dBm power meter sensitivity) are manually adjusted to horizontal $\ket{H}$ and diagonal $\ket{D}$ states referenced to the calibrated polarimeter and switched by a micro-electromechanical systems (MEMS) switch driven by 50\% duty-cycle rectangular pulses at $\sim$10~Hz. As long as the fiber components \emph{inside} each node (Alice, Bob, Charlie) are  taped down and stable, 
the transmitter and receiver can be trusted to maintain self-consistent mutually unbiased reference bases for polarization tracking. (In practice, we found daily recalibration sufficient in our laboratory conditions.)

A Raspberry Pi controls both the MEMS switch and a four-channel electronic polarization controller (EPC; OZ Optics EPC-400), which induces controllable fiber birefringence through piezoelectric actuators. The EPC is updated in real time using a gradient-descent-style routine. At a high level, the approach operates as a greedy coordinate search with an adaptive step size, augmented by stochastic perturbations to escape stagnation, which is similar in spirit to common polarization-tracking routines like golden section search~\cite{Mishra2025} and hill climbing~\cite{Cai2021SAGD}. Mathematically, the system continuously computes the reference fidelity $\cF_\cR$ via \crefrange{eq:fidU}{eq:fractions} and adjusts EPC voltages whenever $\cF_\cR$ drops below a predefined threshold, which we take as $\cF_\text{th}=0.98$ for all experiments below.

In tandem with the classical tracking system, the output of a polarization-entangled photon source (OZ Optics; DTS0187) based on type-II parametric down-conversion in periodically poled silica fiber~\cite{Zhu2012Direct, Zhu2013PoledFiber} is demultiplexed by wavelength: the L-band photon ($\sim$1566-1606~nm) is sent through a polarization analyzer (NuCrypt PA-1000), detected locally by a superconducting nanowire single-photon detector (SNSPD; Quantum Opus), and time-stamped (Swabian TTU-1024); the C-band photon ($\sim$1526--1566~nm) is multiplexed with the classical tracking signal via a 100~GHz dense wavelength-division multiplexing (DWDM) filter centered at 1610.85~nm, transmitted through the deployed fiber channel, demultiplexed and filtered at the receiver using three DWDMs and a 5~THz pulse shaper (WaveShaper 4000A), projected onto the desired polarization state (NuCrypt PA-1000), detected by an InGaAs single-photon avalanche photodiode (SPAD; ID Quantique ID Qube), and time-stamped (Swabian TTU-1024). Timing synchronization is realized via a White Rabbit network  (Orolia) over parallel optical fibers that also carry classical communications between the nodes~\cite{arefur2024AAPT,Alshowkan2024}. Coincidences are defined within a 2~ns window. 


\section{Results}


Arguably, the greatest limitation of AAPT-based channel tracking is speed: because the approach relies on counting coincidences from a comparatively weak entangled-photon source, there is no expectation to match the $\sim$10~Hz update rates achieved by the classical polarization tracking channel. Therefore, the extent to which the total time of each tomography can be reduced sets a critical practical limit for validating the quantum process in real time. To determine this limit empirically, we first focus on $S=16$ total state projections~\cite{james2001measurement}, namely $\ket{\varphi_s}=\ket{\alpha\beta}$ where $\alpha\beta\in\{HH, HV, HD, HL,VH,VV,VD,VL,DH,DV,DD,DR,\\RH,RV,RD,RL\}$. The states $\{\ket{H},\ket{V},\ket{D},\ket{A}\}$ are as defined earlier, and $\ket{R}=\frac{1}{\sqrt{2}}(\ket{H}+\imag\ket{V})$ and $\ket{L}=\frac{1}{\sqrt{2}}(\ket{H}-\imag\ket{V})$. Tomographically complete, this set reduces total acquisition time by 56\% compared to the over complete 36-projector set employed in our previous AAPT experiment~\cite{arefur2024AAPT}.

\begin{figure}[tb!]
\includegraphics[width=\columnwidth]{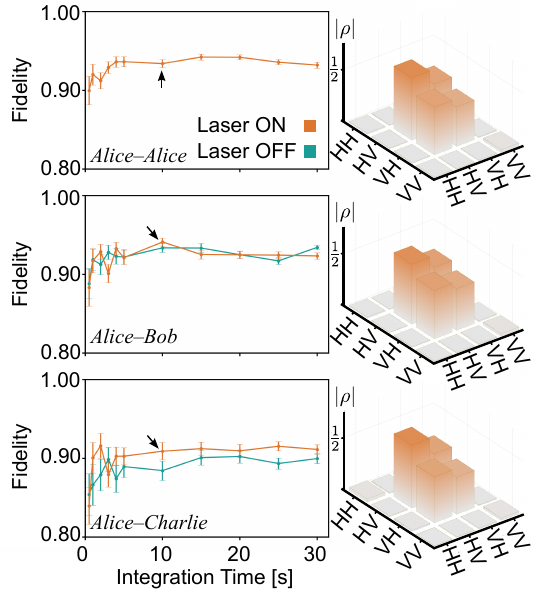}
    \caption{State fidelity measured at various integration times with absolute values of the ``Laser ON'' Bayesian mean density matrices at 10~s integration. The orange line corresponds to the classical tracking sources being on during measurement with the teal representing the classical source being off. The ancilla is measured locally without a multiplexed channel.}
    \label{fig:int-time}
\end{figure}

Next, to choose an integration time over which coincidences for each projector combination are collected, we perform initial two-qubit state tomography tests with (i)~both detectors at Alice, (ii)~one at Alice and one at Bob, and (iii)~one at Alice and one at Charlie, with a range of integration times from 0.5--30~s. The measurement is implemented with both the multiplexed classical source on and off to quantify the impact of the added reference sources to the combined channel. As shown in Fig.~\ref{fig:int-time}, the estimated fidelity has high uncertainty for 0.5--10~s integration time, but then levels off, with
no statistically significant difference for 10--30~s. The plotted Bayesian mean density matrices at 10~s integration (corrected for local rotations~\cite{Grondalski2002, Alshowkan2022c, arefur2024AAPT}) show good agreement with the ideal $\ket{\psi^+}=\frac{1}{\sqrt{2}}(\ket{HV}+\ket{VH})$ Bell state, albeit with some imbalance in the $HV$ and $VH$ contributions. We note that our total detected coincidence rate over all four states in a basis is $\sim$1800~s$^{-1}$ for Alice--Alice, $\sim$1600~s$^{-1}$ for Alice--Bob, and $\sim$780~s$^{-1}$ for Alice--Charlie. For all future measurements, we leverage a 10~s integration time over all 16 projectors, for a total acquisition time of $\sim$163~s per tomography (the extra 3~s stemming from instrument communication and reconfiguration time). Using this measurement method combined with the classical reference described earlier, we monitor the state of the channel across three scenarios.

\begin{figure*}[tb!]\centering
\includegraphics[width=\textwidth]{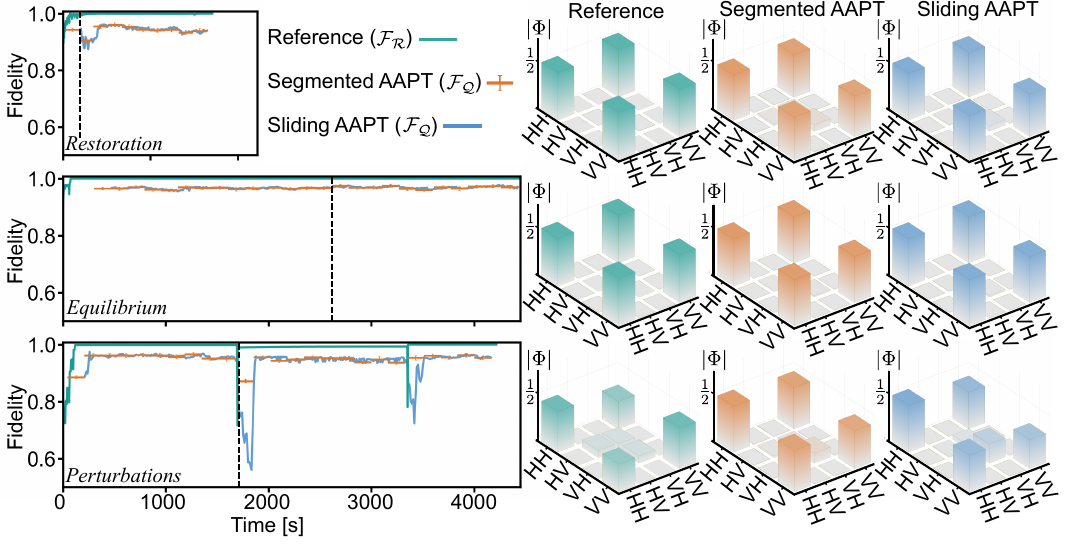}
    \caption{Quantum link verification between Alice and Bob with exemplar Choi matrices.  The teal curves show the reference fidelity $\cF_\cR$ measured from the classical tracking system. The orange symbols with uncertainty bars show the fidelity $\cF_\cQ$ measured through segmented AAPT, in which each tomography is computed from 16 distinct measurements; the width of each symbol corresponds to the total measurement time (163~s). The blue lines show  AAPT computed via a sliding 16-projector window advanced in 10~s steps across the acquisition stream. The Choi matrix magnitudes $|\Phi|$ correspond to the times marked by the dashed black lines, where the segmented AAPT Choi result is either that immediately to the right or intersected by the relevant vertical line. The measured loss of the link is 0.38~dB.}
    \label{fig:A-B}
\end{figure*}

\Cref{fig:A-B} shows the results for the Alice--Bob link. In the top plot, where manual polarization controllers are adjusted to intentionally suppress fidelity prior to optimization, the AAPT measurements confirm rapid attainment of a stable equilibrium, yet with a lower fidelity $\cF_\cQ$ and tracking bandwidth compared to the classical system's $\cF_\cR$ based on high flux and the assumption on unitarity. The middle plot depicts a test in which the optimization algorithm has been allowed to converge before the start of AAPT measurements. In this case, the fidelity remains consistently high across 25 consecutive tomographies, indicating that the channel is stable over extended periods. The roughly constant gap in fidelity between $\cF_\cR\approx0.99$ and $\cF_\cQ\approx 0.97$ can be attributed to the difference in channel unitarity, which is closely related to the purity of the Choi matrix $\Tr\Phi^2$~\cite{Wallman2015}.  Whereas $\Tr\Phi_\cR^2=1$ holds for the reference process always (by specification), $\Tr\hat{\Phi}_\cQ^2\approx 0.94$ for the AAPT estimates. Finally, the bottom plot highlights the difference in temporal sensitivity between the two methods. When abrupt perturbations are introduced by manually adjusting a polarization controller (at 0, 1695, and 3350~s), the optimizer responds rapidly to restore high fidelity, whereas the AAPT fidelity exhibits slower dynamics due to its 163~s measurement time.

Importantly, this 163~s binning can be reduced considerably by adopting a sliding measurement window in which Bayesian AAPT is performed immediately after receiving a new 10~s projection result $N_s$, by combining this with the most recent 15 projections. In this ``sliding AAPT'' approach, successive tomographies share 15 out of 16 measurements in common, and so are no longer independent of each other. Yet they can more readily sense the impact of changes to the count distribution as they appear in the data collection process.
The solid blue lines in \cref{fig:A-B} trace the fidelity inferred by sliding AAPT. In the bottom plot in particular, sliding AAPT ascertains the drops at 1695~s and 3350~s much more accurately than segmented AAPT, which indeed completely misses the 3350~s dip. Sliding AAPT still takes a relatively long time to validate the tracking system's return to a high-fidelity channel, which makes sense given its reliance, like segmented AAPT, on data extending 163~s into the past. However, its opportunistic use of the most recent datasets at any give time imparts greater agility and responsiveness to channel fluctuations closer to that of the classical tracking system.

The right side of \cref{fig:A-B} shows exemplar Choi matrices, which provide a visual representation of the differences in each method at a given time slice. Note that the ideal identity channel $U=\mathbbm{1}$ corresponds to $\Phi=\ket{\phi^+}\bra{\phi^+}$ with $\ket{\phi^+}=\frac{1}{\sqrt{2}}(\ket{HH}+\ket{VV})$---i.e., equal-amplitude peaks on the four corners. In low-fidelity cases, the matrices exhibit strong weights away from the ideal identity block---often enhanced in the $VH$ component---suggesting some level of polarization-flip errors rather than purely isotropic depolarization~\cite{johnston2011choi,dur2005depolarization}.

As with the state tomography results in \cref{fig:int-time} and for ease of visual representation, we remove
the residual gauge freedom associated with local unitaries in postprocessing. 
Specifically, for each time trace, we select a single unitary operator $V$ that maximizes fidelity of the final point in the plot with with respect to $\ket{\phi^+}\bra{\phi^+}$, and then adjust all raw Choi matrices to $(\mathbbm{1}\otimes V)\Phi_\cQ (\mathbbm{1}\otimes V^\dagger)$. We emphasize that this unitary correction is solely for illustrative purposes, with no impact on  system operation or numerical calculations: both the classical tracking system and AAPT operate directly on their respective datasets with no reference to an external unitary.

\begin{figure*}[!tb]\centering
\includegraphics[width=\textwidth]{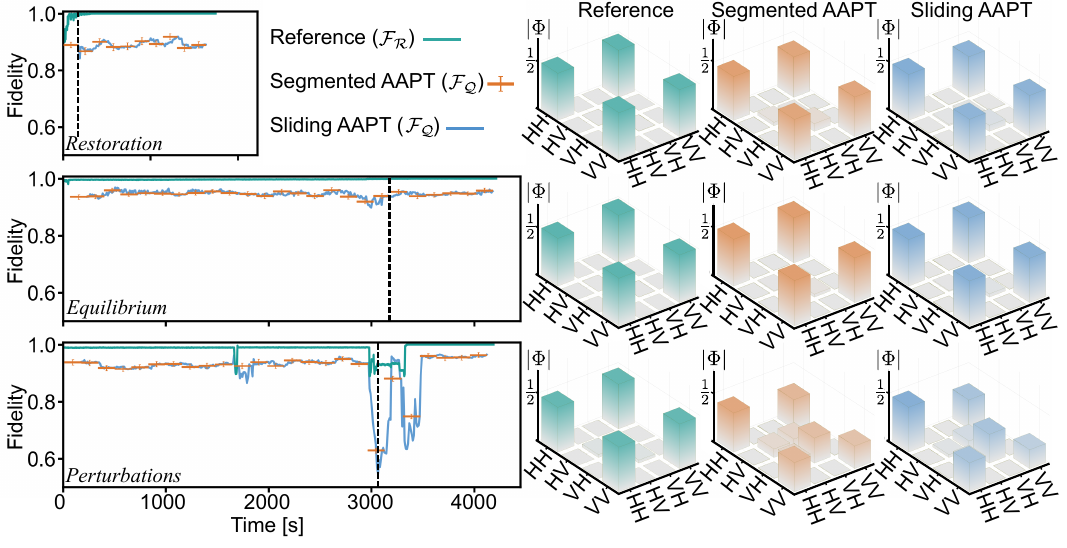}
    \caption{Quantum link verification between Alice and Charlie with highlighted Choi matrices. The teal lines show the fidelity $\cF_\cR$ measured from the classical polarization references. The orange symbols with uncertainty bars show the fidelity $\cF_\cQ$ measured through AAPT with the length corresponding to the total measurement time (163 seconds). The blue lines show AAPT computed via a sliding 16-projector window advanced in 10~s steps across the acquisition stream. Example Choi matrices correspond to the times marked by the dashed black line, with the segmented AAPT result denoting the symbol immediately to the right or in the middle of the line. Here the measured loss of the link is 4.02 dB.}
    \label{fig:A-C}
\end{figure*}

\Cref{fig:A-C} summarizes the experiments performed across the Alice--Charlie link, which endures a higher $4.02$~dB link loss (primarily from poor splicing connections in Charlie). Here we apply perturbations at 0, 1678, and 2981~s using the same method as before. As a whole, the behavior is very similar to that of \cref{fig:A-B}, apart from slightly lower AAPT fidelity values of $\cF_\cQ\approx 0.95$ compared to $\cF_\cQ\approx 0.97$ for Alice--Bob, which can be attributed to the higher loss (and hence lower signal-to-noise ratio). 
As in \cref{fig:A-B}, the stair-step appearance of the converged reference fidelity $\cF_\cR$ results from the threshold $\cF_\text{th}=0.98$, for the algorithm is designed to stop at any fixed value $\cF_\cR\in[\cF_\text{th},1]$. For example, the compensation setting found at $\sim$1700~s in the bottom plot of \cref{fig:A-C} happens to be lower in this allowed range than that found at $\sim$3300~s. 

\section{Discussion}
\label{sec:discussion}
Overall, the consistency between the reference and AAPT fidelities across the tests in \cref{fig:A-B,fig:A-C} confirm the viability of our approach for in situ monitoring of time-dependent quantum links. In light of the growing importance and adoption of polarization tracking methods in deployed quantum testbeds, this verification tool provides a window  into the quantum performance of a channel otherwise probed only classically, opening the door to further experimentation with multiplexed links in which additional quantum and classical channels share the fiber lightpath with both the tracking system and AAPT. In this vein, the coarse wavelength division multiplexing of the current experiment---i.e., quantum light in the C-band, classical in the L-band---could be complemented by additional multiplexing strategies, such as DWDM and time-division multiplexing (TDM). Indeed, TDM offers a particularly practical path for quantum--classical coexistence and is already widely employed for polarization tracking~\cite{kucera2024quantum,treiber2009-8,craddock2024automated17}. Our approach is directly compatible with \emph{slow} TDM~\cite{lukens2025hybrid}, in which long quantum and classical frames are interleaved in time, and so it would be useful to explore the extent to which our method could perform even better under TDM due to reduced concurrent quantum--classical crosstalk.

Another interesting, though more exploratory, direction concerns space-division multiplexing with multicore fibers~\cite{Puttnam2021,PRXQuantum2021}, in which the quantum channel may occupy one core while classical data are carried in another. An important question in this regime is how faithfully polarization transformations in one core track those in neighboring ones. Preliminary studies have shown that polarization drift between cores can be highly correlated under common perturbations~\cite{Mazur2015}, and theoretical models of intercore crosstalk including polarization-mode coupling can provide further insight~\cite{Antonelli2012}. However, a comprehensive, time-resolved characterization of polarization correlations across cores remains an intriguing subject for future work~\cite{lukens2025hybrid}.

To further improve upon this experiment, the AAPT integration time could potentially be reduced by employing a brighter entangled photon source, such as the periodically poled lithium niobate Sagnac design of Ref.~\cite{Alshowkan2022c}, which we estimate to output $\sim$300$\times$ higher flux over the C+L-band than the periodically poled silica fiber design here~\cite{Zhu2012Direct}. With our ``sliding AAPT'' approach already providing a means of continuous process monitoring via overlapping tomographic windows, a brighter source would be able to shorten these sliding windows even further, possibly refining the temporal resolution for which sliding AAPT can track a channel down to subsecond levels.

At such short integration times, the computational cost of MCMC will likely become the main bottleneck. In our current proof-of-principle demonstration, MCMC was performed offline, and no effort was made to optimize the postprocessing pipeline. As a conservative setting, we performed $2^{22}$ mutation steps per chain, amounting to about five minutes per tomography. However, such long chains could likely be reduced considerably by leveraging adaptive stopping rules based on convergence diagnostics (such as the Gelman--Rubin statistic ~\cite{GelmanRubin1992MCMC, Vehtari2021ImprovedRhat}). On the systems side, asynchronous execution and consolidated storage formats could reduce communication overhead, while lighter numerical routines (e.g., selective single-precision operations or GPU batching) could further shrink per-iteration costs. Additionally, based on the parallel pCN MCMC approach reported in Ref.~\cite{Nguyen2025}, we predict $\sim$100-fold computational speedups would be attainable through parallelization. Accordingly, we  see no major roadblocks to reducing the MCMC wall clock time to below whatever integration time may be selected in future Bayesian AAPT implementations.

In summary, our results demonstrate that deployed links can be classically compensated and quantum-characterized through in situ AAPT. While classical monitoring approaches more sophisticated than the unitary model adopted here can detect depolarization, they still lack sensitivity to effects present at the low-flux quantum level, thus underscoring the unique and persistent advantages of our quantum-based method, whose nonintrusive nature and compatibility with ongoing channel use should make it a valuable diagnostic tool for emerging quantum--classical coexistence networks.

\begin{Center}
{\centering\textbf{ACKNOWLEDGMENTS}}
\end{Center}

We thank Arizona State University students James Crandall, Jenifer Sherrill, Raihan Nishat Tonmoy, Zachary Oliver, and Delaney Carrigan for discussions related to earlier versions of the experiment.
This work was performed in part at Oak Ridge National Laboratory, operated by UT-Battelle for the U.S. Department of Energy under Contract No. DE-AC05-00OR22725. Funding was provided by the Laboratory Directed Research and Development program of Sandia National Laboratories (EPIQ, APT) and the National Science Foundation (Research Experiences for Undergraduates Program).

\bibliography{references}
\end{document}